\documentstyle[preprint,prb,aps,epsf]{revtex}
\def\inseps#1#2{\def\epsfsize##1##2{#2##1} \centerline{\epsfbox{#1}}}

\begin{document}
\draft
\title{DECIPHERING THE FOLDING KINETICS OF TRANSMEMBRANE HELICAL PROTEINS}
\author{Enzo Orlandini$^{1}$, Flavio Seno$^{1}$, Jayanth R. Banavar$^{2}$,
Alessandro Laio$^{3}$ \& Amos Maritan $^{3,4}$}
\address{$^{1}$INFM-Dipartimento di Fisica, Universit\`a di Padova, Via\\
Marzolo 8, 35131 Padova, Italy\\
$^{2}$ Department of Physics and Center for Materials Physics, The\\
Pennsylvania State University, University Park, 16802 Pennsylvania\\
$^{3}$ INFM - International School for Advanced Studies, Via Beirut 4, 34014 
\\
Trieste, Italy \\
$^{4}$ The Abdus Salam ICTP, Strada Costieria 11, 34100 Trieste, Italy }
\date{\today }
\maketitle

\begin{abstract}
Nearly a quarter of genomic sequences and almost half of all receptors that
are likely to be targets for drug design \cite{editorial} are integral
membrane proteins. Understanding the detailed mechanisms of the folding of
membrane proteins is a largely unsolved, key problem in structural biology.
Here, we introduce a general model
and use computer simulations to study the equilibrium properties and
the folding kinetics of a $C_{\alpha}$-based two helix bundle fragment
(comprised of $66$ amino-acids) of Bacteriorhodopsin. 
Various intermediates are identified and their free energy are
calculated
toghether with the free energy barrier between them.
In $40 \% $ of folding trajectories, 
the folding rate is considerably
increased by the presence of non-obligatory intermediates acting as traps.
In all cases, a substantial portion of the helices is rapidly formed. This
initial stage is followed by a long period of consolidation of the helices
accompanied by their correct packing within the membrane. Our results
provide the framework for understanding the variety of folding pathways of
helical transmembrane proteins.
\end{abstract}

%\input{psfig}
%\psdraft
%\psfullacs insert.txt

\vskip 0.3cm

\vskip -1truecm

\newpage
Considerable effort has been expended to understand the dynamics of the
folding and biological functionality of proteins. Whereas the behavior of
small water soluble globular proteins is reasonably well understood both
experimentally and theoretically \cite{Fersht3,Karplus}, much less is known
about membrane 
proteins (MP) \cite{white1,OM97,vH96,booth} that cross biological
membranes. Transmembrane proteins (TMP) are the most important and best
studied class of MP \cite{white1,OM97,Biggin}. They are characterized by the
presence in their primary structure of long segments ($20 - 30$) of amino
acids with a high degree of hydrophobicity. In the native structure, these
correspond to the transmembrane segments which are inserted in the lipidic
interior of the membrane \cite{goto}. These segments are predominantly made
up of $\alpha$-helices and $\beta$-sheets.
The stability of $\alpha$-helices and $\beta$-sheets inside the membrane
follow from the formation of hydrogen bonds between the backbone atoms --
other possibilities are excluded within the apolar enviroment 
\cite{white1,PE90}.

Phenomenological models have proved to be powerful for interpreting
experimental data. The most common of these is the {\em Two-Stage model}
based on experimental evidence that the folding of TMP occurs in two stages.
In the first stage, $\alpha$-helices and $\beta$-sheets are formed with the
full native state structure being formed in a distinct second stage \cite
{PE90}. A more refined model\cite{white1}
 takes into account four main steps:
partitioning, folding, insertion and association. Recently, Pappu et al. 
\cite{Papu} have used a potential smoothing algorithm to predict
transmembrane helix packing in good accord with experimental data.

Milik and Skolnick \cite{MS1,MS2} have carried out careful Monte Carlo
studies of the insertion of peptide chains into lipid membranes and have
proposed a new hydropathy scale based on experimental data obtained by
studying the interactions of tripeptides with phospholipid membranes\cite
{JW87} and the self-solvation effect in protein systems\cite{Rose88}. 
Recently Wimley and White\cite{preprint} have designed transmembrane 
peptides that spontaneously insert across bilayers but yet have
measurable monomeric water stability, opening the way for the determination
of the thermodynamic cost of partitioning  hydrogen bonded peptide bonds
into the membrane hydrocarbon core.

The Monte-Carlo results of Milik and Skolnick\cite{MS1,MS2}
 are in good accord with Engelman and Steitz's helical
hairpin hypothesis\cite{ES81} further extended by Jacobs and White\cite
{JW89,JW86}. The unfolded chain is first adsorbed onto the membrane
interface, driven mostly by the hydrophobic effect and electrostatic
lipid-protein interactions \cite{one,two,three}. A polypeptide chain has a
greater possibility, while anchored to the interface, of saturating its
internal hydrogen bonds and forming helices. Such helical fragments have a
greater propensity to subsequently diffuse into the lipid phase.

A detailed study of TMP has not yet been possible because little is known
about the interactions between amino acids inside the membrane or between
them and the lipid molecules. Here, we adopt a simple, yet powerful,
strategy for attacking the folding properties of TMP that circumvents
this shortcoming. Our novel approach is based on extensive studies 
of the folding of globular proteins which have underscored
the important role played by the topology of the native state in
controlling both the functionality and the main features of the folding
process.
Nature uses a rich repertory of twenty
kinds of amino acids with sometimes major and at other times subtle
differences in their interactions with the solvent and with each other in
order to design sequences that fit the putative native state with minimal
frustration \cite{Woly}. Thus a fruitful and general strategy for the study of
protein folding would be to extract information on the folding process
directly from the topology of the native state. 

Our study here focuses on the
folding process by using a  tractable approach (described in the Methods
Section) that by-passes the details of the complex interactions of the protein
in the lipid enviroment by introducing effective potentials, induced by the
presence of the membrane and the associated interface region, that stabilize
the native state structure. 
The validity of the approach based on the native state topology, in the
case of globular proteins, has been confirmed a posteriori from the agreement
between theory and experimental findings 
\cite{Fersht3,Fersht2,Fersht1,Micheletti2,Eaton1,Finkel,Baker,7,8,9}. 
The  approach proposed here is similar in spirit and ought to be 
a tool and a guide for the difficult experimental situation of
TMP\cite{preprint}. 
Our model allows a complete characterization of the thermodynamics and the
dynamics  of the full folding process.  
Due to the small number of degrees of freedom involved, the dynamics of the
system can be simulated for the full folding process. Moreover, the free
energies of the most relevant intermediate states and free energy profiles 
along the reaction paths connecting them can be explicitly calculated by
thermodynamic integration (see Methods).
Thus the model is able to quantitatively
discriminate between the possible reaction paths envisaged for the insertion
process of TMP across the membrane\cite{white1}, a feature that is
not an obvious consequence of the structure of the model.
Where there is overlap, our model captures the qualitative features of the
earlier simulations of  Milik and Skolnick\cite{MS1,MS2}.

The TMP we considered is made up of the first 66 amino acids of
bacteriorhodopsin consisting of two $\alpha $-helices (Fig. $1a$).
It has been shown that the first two helices of 
bacteriorhodopsin can be considered as
independent folding domains\cite{KSE92}. Furthermore, the side-by-side
interactions between transmembrane helices play a key role in the
stabilization of the protein structure\cite{KE92}.

Our studies were carried out using a Monte Carlo algorithm that has been
proven to be extremely efficient for interacting hetero-polymers
(Methods). The behaviour of the structural similarity between the system
equilibrated at temperature $T$ (measured in dimensionless units) and the
native state is shown in Figure $1b$ in terms of the average fraction of
native state contacts as a function of $T$ and partitioned depending on
their positions with respect to the membrane. The three curves correspond
respectively to the average fraction of native contacts inside ($q_m$) ,
outside ($q_b$) and across ($q_s$) the membrane (see Methods). All these
curves, well separated at high T, collapse for $T$ below the transition
temperature $T_C \sim 0.6$, indicating a cooperative effect in the folding.
On monitoring the free  energy as a function of the energy around $T_C$,
one observes additional local minima (besides those corresponding to
the unfolded and folded states) suggesting the presence of an intermediate.

The intermediate is
characterized by having the two helices almost completely formed but not yet
correctly inserted across the membrane.
A metastable state in which the protein exists at the membrane interface
ought to be expected on general grounds. Indeed a
generic heteropolymer with hydrophobic and hydrophilic aminoacids, of which a
TMP is a particular case, has a favorable conformation
which is  localized near surfaces
between two selective media (the outside and the inside part of the membrane
in the present case)\cite{Garel,Riva}.  At not too high temperatures, the
gain in energy to place hydrophobic/hydrophilic protein segments in their
preferred enviroment compensates the  entropy loss for being localized at the
interface with respect to remain in the bulk phase. Thus, even though our model
does not explicitly contain information on the character of the amino acids,
it is able to predict this feature.
 
The presence of these extra minima  suggests that non-constitutive
membrane proteins would fold with multi-state kinetics corresponding to
on-pathway intermediates. To establish their nature of  and their influence on
the dominant folding pathways, we have performed a detailed analysis of the
folding kinetics. Each independent kinetic folding simulation was started with
the equilibrated denaturated state at $T^* = 2.5$ . The protein is placed
initially outside the membrane in the interface region \cite{white1}, at a
distance comparable to the average size of the denatured protein and then
suddenly quenched to a temperature ($T=0.4$) well below the transition
temperature. This case simulates the folding kinetics of non-constitutive
membrane proteins, i.e. proteins that do not need a translocon providing a
'tunnel' through which the protein is injected into the lipid bilayer. Folding
to the native state occurs mainly through the states depicted in Figure $2a$
with the dominant pathways shown in Figure $ 2b $.

In all the pathways, the system goes from the unfolded state, $U$ to state $
HI$ in which $80\%$ of the secondary structure is formed (see $q$ in Figure $
3c$) and disposed horizontally along the interface. The free energy of this
state (measured with respect to the free energy of the fully 
folded state) is $\sim
2.4$ T$_C$. This state corresponds to the formation of around $70\ \%$ of the
membrane contacts. The average time $\tau _{HI}$ to reach state $HI$ is of
the order of $500$ Monte Carlo steps (see Figures $3$ and $4$;  
each Monte-Carlo step
corresponds to 50000 attempted local deformations.). State HI
turns out to be an obligatory on-pathway intermediate of the folding
kinetics for non-constitutive MP in agreement with the general
argument mentioned above. Once the protein reaches state $HI$, it
undergoes a relatively slow process of self-arrangement in order to insert and
assemble the secondary structures across the membrane. This process is the
rate-limiting step of the folding process, since it involves the
translocation, through the lipidic layer, of a substantial number of
hydrophilic residues. Among the possible pathways, starting from $HI$, the
most frequent ($60\%$ of the cases) and the fastest turn out to be $
U\rightarrow HI\rightarrow HV\rightarrow N$. A quantitative characterization
of this dominant pathway is presented in Figures 3 (for a single folding
process) and 4 (as an average over $40$ folding processes). The intermediate 
$HV$ is characterized by having one $\alpha $ helix inserted across the
membrane and is reached in an average period corresponding to a significant
fraction of the total folding time (see Figure $3$). The free energy in this
state is $\sim 0.98$ $T_C.$ The  free energy barrier between $HI $ and $HV$ is
at $\sim $ $4.31$ $T_C$ (hence, the rate constant of the transition
$HI\rightarrow HV$ is proportional to $k_{HI\rightarrow HV}=\exp \left(
-\left( 4.31-2.4\right)T_C/T \right) $). The full free energy profile
versus a reaction coordinate is shown in Fig. 5. 
The last part of the folding process corresponds to the insertion of the
second helix and the assembly of the two secondary structures into the
native state structure. This process lasts approximately one third of the
folding time along the pathway $U\rightarrow HI\rightarrow HV\rightarrow N$.
The quasistatic free energy barrier between $HV$ and the folded state is $
\sim $ $1.66$ $T_C$. The rate costant of the transition 
$HV\rightarrow N$
is, therefore, proportional to 
$\exp \left( -\left( 1.66-0.98\right)T_C/T\right)$. 
These results are consistent
with the time scales observed in the unconstrained folding dynamics. 
At the end, the protein is completely packed, ($q_m$ saturates to 1 (Figures 
$3a$ and $4a$)) and the helices are correctly positioned across the membrane
(note the second jump in the $z$ coordinate of the center of mass in Figures 
$3b$ and $4b$).

Much slower dynamics can occur when non-obligatory intermediates are visited
by the system. These long lived states ($\{ I \}$ in Figure $2a$)
involve a distribution of misfolded regions that trap the system and are
characterized by having most of the inter-helical contacts formed (assembly
of the secondary structures) but with the two ${\alpha }$-helices still
incorrectly positioned. Note, for example, that in states $\{ I \}$ , only
transmembrane contacts and some contacts outside the membrane are misplaced
and they account for only a small fraction of the native state energy. For
this reason, in the states $\{I \} $, the 
free energy is $\sim $ $1.44$ $T_C$, only
slightly higher than the free energy of $HV$. The folding can proceed
from 
$\{I \}$ either by disentangling the two helices and passing through the
obligatory intermediate $HV$, or by the simultaneous translocation through
the membrane of the two helices. These processes, however, entail the
crossing of a big free energy barrier ($\sim $ $5.18$ $T_C$ for the first
process   and $6.1$ for the second)
and  happen with low probability. Indeed, at sufficiently low
temperatures, the loss in energy of the interhelical contacts is not
compensated by the gain in the configurational entropy due to the uncoupling
of the ${\alpha -}$helices. Thus below the folding temperature, I-states act
as trapping regions for the system and when trapped, the protein spends most
of the time during folding in this state.

In summary, we have presented detailed calculations of helical transmembrane
proteins leading to a vivid picture of the folding process. Our strategy
relies on the dominant role played by the topology of the native state
structure and by the effective geometry imposed by the membrane
and provides a picture which would be expected to be quite accurate for
well-designed sequences that are a good fit to the target native state
conformation.
It is interesting to note that , with our choice of the
parameters,
the pathway in which the helices assemble outside the membrane and are
inserted later is unlikely to occur.

Models based on the topology of the native state structure have been
remarkably successful \cite{Micheletti2,Eaton1,Finkel,Baker}
in correctly describing the main features 
of the folding process determined in experiments
\cite{Fersht3,Fersht2,Fersht1,7,8,9} for various globular proteins. A similar
approach has been generalized here to the almost virgin field of
transmembrane proteins where experiments are rather
difficult\cite{preprint,white1,booth}. 
Our findings do not depend on the precise values of the
$\epsilon$
parameters introduced in the model underscoring the robustness of the
results.
Our approach predicts
a folding process involving multiple pathways with a dominant folding channel.
The simpliciy of our model allows for a quantitative description of
all
the pathways since we can monitor the correct/uncorrect formation
of native contacts and compute free energy profiles. Further details
not captured by the present
approach arising from  amino-acid specific interactions among
themselves, with the solvent and in particular with the interior of
the membrane may of course
change the quantitative nature of the results. 
However,  our model, which captures the bare essentials of
a membrane protein, ought to provide a zeroth order
picture of the folding process. Also, as experimental data becomes
available,
the results could be benchmarked with models of this type to glean
the other factors that matter.

\section{Methods}

We represent the residues of the membrane protein as single beads centered
in their $C_{\alpha}$ positions. Adjacent beads are tethered together into a
polymer chain by a harmonic potential with the average $C_{\alpha}-C_{%
\alpha} $ distance along the chain equal to $3.8 {\AA}$. The membrane is
described simply by a slab of width $w = z_{{\rm max}}-z_{{\rm min}}=
26 {\AA}$. Two
non-bonded residues $(i,j)$ form a contact if their distance is less then $%
6.5 {\AA}$. In the study of globular proteins, the topology of the native
state is encoded in the contact map giving the pairs $(i,j)$ of non-bonded
residues that are in contact. Here, in addition, the locations of such pairs
with respect to the membrane becomes crucial. The contacts are divided into
three classes: {\sl membrane contacts} where both $i$ and $j$ residues are
inside the membrane, {\sl interface contacts} with $i$ and $j$ in the
interface region \cite{white1} outside the
membrane and {\sl surface contacts} with one residue inside the membrane and
the other outside. Thus a given protein conformation can have a native
contact but improperly placed with respect to the membrane ({\sl misplaced
native contact}). The crucial interaction potential between non-bonded
residues $(i,j)$ is taken to be a modified Lennard-Jones 12-10 potential: 
\begin{equation}
\Gamma(i,j) \left [ 5\left ( \frac{d_{ij}}{r_{ij}}\right)^{12}- 6\left( 
\frac{d_{ij}}{r_{ij}}\right)^{10}\right ] +5\ \Gamma_1(i,j)\left ( \frac{%
d_{ij}}{r_{ij}}\right)^{12}.
\end{equation}
The matrices $\Gamma(i,j)$ and $\Gamma_1(i,j)$ encode the topology of the
TMP in the following way: if $(i,j)$ is not a contact in the native state $%
\Gamma(i,j)=0,\Gamma_1(i,j)=1$; if $(i,j)$ is a contact in the native state
but not at the proper location (i.e. a misplaced contact) $%
\Gamma(i,j)=\epsilon_1,\Gamma_1(i,j)=0$; if $(i,j)$ is a native state
contact in the proper region $\Gamma(i,j)=\epsilon,\ \Gamma_1(i,j)=0$. This
model is intended to describe the folding process in the interface and in
the membrane region. Our interaction potential (similar in spirit to a well
known model\cite{Go} for globular proteins (see also other approaches 
that model helix formation \cite{new1,new2})) assigns two values to the energy
associated with the formation of a native contact, 
$\epsilon$ and $\epsilon_1 $. 

%They take into account, in an effective way, the propensity to
%form the correct contacts in the correct region (inside, outside or across
%the membrane) corresponding to the native state structure.
%

The model captures the tendency to form native contacts.  In addition,
in order to account for the effective interactions between the membrane
and the protein, the model assigns a lower energy, $-\epsilon$, to the
contact which occurs in the same region as in the native state structure 
compared to 
$-\epsilon_1$ when the contact is formed but in the wrong region of space.
This feature proves to be crucial
in determining the mechanism of insertion of the protein across the
membrane in order to place all native contacts in the same regions as in the
native state. Even though the interaction potential is simple
and intuitively appealing, it is not possible to simply guess (without
detailed calculations) the
folding mechanism and  quantitatively determine the probability
of occurrence of the 
various folding pathways\cite{white1}.  

When $\epsilon=\epsilon_1$, the protein does not recognize the
presence of the interface-membrane region and the full rotational symmetry
is restored (the system behaves like a globular protein). The
difference in the parameters ($\epsilon-\epsilon_1$)
controls the amount of tertiary
structure formation outside the membrane.  When the difference is small,
the protein assembles almost completely outside the membrane and  the
insertion process would be diffusion limited.
Our results are independent of the
precise values of the energy parameters $\epsilon$ and $%
\epsilon_1$
($\epsilon > \epsilon_1$) as long as they are not too close to each other.

We report here the results of  simulations with $% 
\epsilon_1 = 0.1$ and $\epsilon = 1$.
$r_{ij}$ and $d_{ij}$ are the distance between the two residues $(i,j)$ and
their distance in the native configuration, respectively. In order to account
for the chirality of the TMP, a potential for the pseudodihedral angle
$\alpha_i$ between the $% 
C_{\alpha}$ atoms in a helix corresponding to four
successive locations is added which biases the helices to be in their native
state structure.

The thermodynamics and the kinetics of the model were studied by a Monte
Carlo method for polymer chains allowing for local deformations. The
efficiency of the program (usually low for continuum calculations) has been
increased by full use of the link cell technique \cite{Binder} and by the
multiple Markov chain method, a new sampling scheme, which has been proven
to be particulary efficient in exploring the low temperature phase diagram
for polymers \cite{TJOW}. In our simulation $20$ different temperatures
ranging from $T=2$ to $T=0.17$ have been studied. The free energy is
calculated by reweighting the different temperatures with the
Ferrenberg-Swendsen \cite{FS89} algorithm.

The free energy difference ${\cal F}_B-{\cal F}_A$ between two states A and
B has been estimated as the reversible work that has to be done in order to
go from A to B. Hence, denoting by ${\bf x}\left( \lambda \right) $ a
reaction coordinate connecting A and B (for $\lambda =0$ and $\lambda =1$
the system is in A and B respectively), and by $\left\langle {\bf \cdot }
\right\rangle _\lambda =\left\langle \delta \left( {\bf x}-{\bf x}\left(
\lambda \right) \right) \,{\bf \cdot \,}\right\rangle $, the canonical average
at fixed reaction coordinate, 
\begin{equation}
{\cal F}_B-{\cal F}_A=
\int_0^1 d\lambda \,\left\langle {\bf F}\right\rangle
_\lambda \cdot \frac{d{\bf x}\left( \lambda \right) }{d\lambda }\simeq
\sum_i
\left. \left\langle {\bf F} \right\rangle _\lambda  \cdot
\frac{d{\bf x}\left( \lambda \right)} {d\lambda }
\right| _ {\lambda =\frac{\lambda _i+\lambda _{i+1}}2} \, \,
\left( \lambda _{i+1}-\lambda _i\right)
\end{equation}
where ${\bf F}$ is the force and $\left\{ \lambda _i,i=1,\dots \right\} $ is
a suitably dense partition of the interval $\left( 0,1\right) $. The average value $%
\left\langle {\bf F}\right\rangle _{\lambda _i}$ at each $\lambda _i$ is
computed by a long (more than 5000 steps) Monte Carlo run performed with
dynamics satisfying the constraint ${\bf x}={\bf x}\left( \lambda _i\right) $
. The free energy differences obtained with this method are accurate to within $
\sim $ 0.1$\,T_C$ for the various states whereas the free energy
barriers are accurate within $\sim $ 0.5$\,T_C$ . 
This error takes into account possible hysteresis effects
due to the finite simulation time.

\vskip 1truecm

\noindent {Acknowldgements}

\noindent {We thank Cristian Micheletti for fruitful discussions and
Steve  White for a critical reading of the manuscript
and for many enlightening suggestions. This work
was supported by INFM (PAIS project), MURST-COFIN99,
NASA and the Donors of The Petroleum Research Fund administered
by The American Physical Society. }

\vskip 1truecm

%%%%%% FIGURE
\newpage

\section{Figure legends}

{\bf Figure 1:} Structure and thermodynamics of the helical transmembrane
protein.\\{\bf a)} Ribbon representation of the two-helix fragment of
bacteriorhodopsin formed by the first $66$ amino-acids. The part inside the
membrane (determined by using the neural network learning algorithm
available at http://www.embl-heidelberg.de/Services/sander/predictprotein/)
is shown in red, the part above (below) the membrane in blue (green). {\bf b)%
} Average equilibrium fraction of native contacts outside, $q_b$ ($\circ$),
inside, $q_m$ ($\Box$), and across, $q_s$ ($\triangle$), the membrane as a
function of the temperature $T$. All these quantities are expressed in
energy unit of $\epsilon$ (see Methods). The folding transition temperature $%
T_C$ when all the curves cross the value $1/2$ is around 0.6. This value is
in accord with the temperature of the heat capacity maxima.

{\bf Figure 2:} Schematic representation of states encountered by
non-constitutive proteins during the folding process.\\In {\bf a)} the red
cylinders denote $\alpha$-helices that reside within the membrane in the
native state. The region inside the membrane is in turquoise whereas the rest
represents the interface region \cite{white1} in which the folding process
starts. State $U$ denotes the denatured state of the protein, $HO$ is a
state in which the helices have been formed but are not yet inside the
membrane whereas $HI$ corresponds to a similar state but with the helices
completely embedded in the membrane without any inter-helical contacts.
Usually the helices form and enter into the membrane separately. $HV$
denotes an obligatory intermediate and $N$ depicts the native state.
The state $\{ I \}$ represents  an ensemble of long lived
conformations in which helices are formed inside the membrane with
several
inter-helical contacts, but with the two $\alpha$-helices still
incorrectly
positioned. This conformations differ in term of packing efficiency of
the helices.
The state $\{ I\}$ is not obligatory for the folding kinetics.
In {\bf b)} the schematic pathways to the native state are shown. In the
most directed path, the
entropy decreases on going from $U$ to $N$. From $HI$ to $HV$
the entropy loss of one helix is not compensated by a corresponding energy
gain until both helices become vertical. This is the principal origin of the
high free energy barrier between the state $HI$ and the native state.

{\bf Figure 3:} Typical time dependence of different parameters as a
function of the Monte-Carlo steps for the pathway $U \rightarrow HI
\rightarrow HV \rightarrow N$.
Fraction of native contacts inside the
membrane ({\bf a}), normalized z-coordinate of the center of mass of the
protein (with respect to that of the native state conformation) ({\bf b}) and
overall fraction of native helical contacts ({\bf c}). Each Monte-Carlo step
corresponds to 50000 attempted local deformations. The transition from state 
$HI$ to state $HV$ is signalled by a sharp jump of the position of the
center of mass. Note that there is no perceptible sign of this transition in
terms of newly formed native contacts. Most of the helical contacts are
formed in the early stages of the folding. This fraction does not
significantly increase until helices correctly assemble and the
inter-helical contacts are formed. The $HV \to N$ transition is reached by a
progressive zippering of the horizontal and vertical helices. This
zippering is usually very quick (few MC steps) and is only slightly slowed
down (see the plateau corresponding to $q_m \sim 0.9$ in {\bf a}) when the
trajectory passes through somewhat deformed conformations. ({\bf d}) Protein
conformations at different times during the folding. The colours red, green
and blue have the same significance as in Figure 1a with the grey bonds
being ones crossing the membrane.

{\bf Figure 4:} Distribution of the fraction of native contacts inside the
membrane\\({\bf a}) and of the normalized $z$-coordinate of the center of
mass ($Rz = \frac{z^{cm}}{z^{cm}_{nat}}$) ({\bf b}). The data were obtained
using 40 independent kinetic simulations with pathway $U \rightarrow HI
\rightarrow HV \rightarrow N$. The grey scale distribution indicates the
probabilities at given times: darker points denote higher probability.

{\bf Figure 5:} 
Free energy profiles along three reaction coordinates at T=0.85\thinspace 
T$_C$. The continuous lines are spline fits to the free energy data
(crosses) . To obtain free energy differences between two states
we estimate the reversible work that has be done to go from one
state to the other. For this purpose, we fix the
z coordinate of a specific residue in order to 
compute the canonical average of the force and then apply eq. (2) 
(See Methods).
The free energy of the native state is defined to be equal to 0. 
({\bf a}) Free energy as a function of the  $z$-coordinate of the 58-th residue
($z=0$ corresponds to the middle of the membrane) starting from HV; 
this forces the second helix to cross the membrane as the protein goes
from HV to N;
the local minimum at z$\simeq $20 corresponds to a state topologically 
equivalent to HV, with the helix containing the 58-th residue fully formed 
on the membrane interface but without any contact with the first helix 
(in HV some of the inter-helices contacts are already formed); ({\bf b}) the 
5-th residue is translocated across the membrane with the
protein starting from state HI and proceeding  
to HV; 
({\bf c}) 
the same as in ({\bf b}), but the initial state is I (see Fig. 2-a)

\begin{figure}[tbp]
\inseps{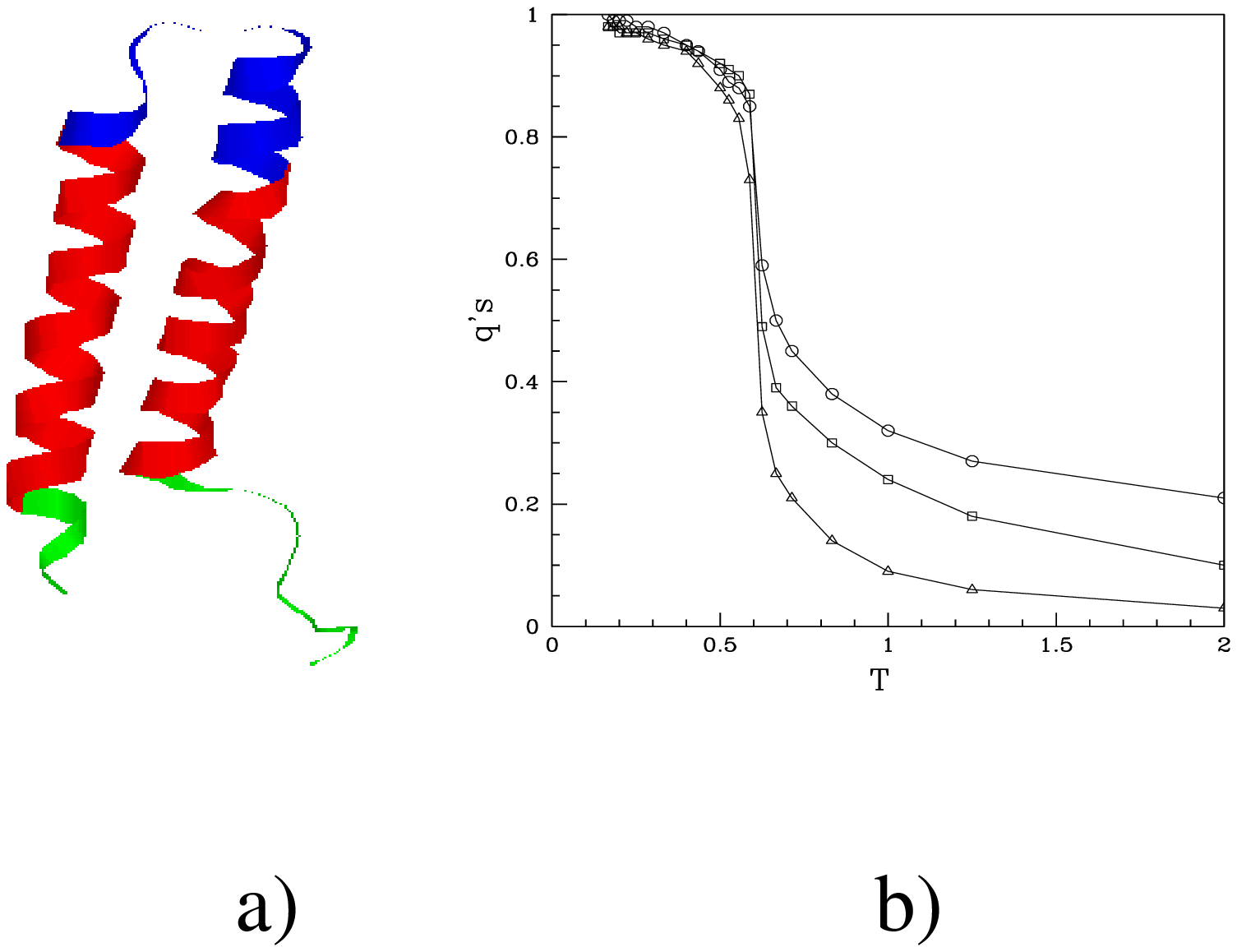}{0.7}
\caption{Seno}
\label{fig1}
\end{figure}

\begin{figure}[tbp]
\inseps{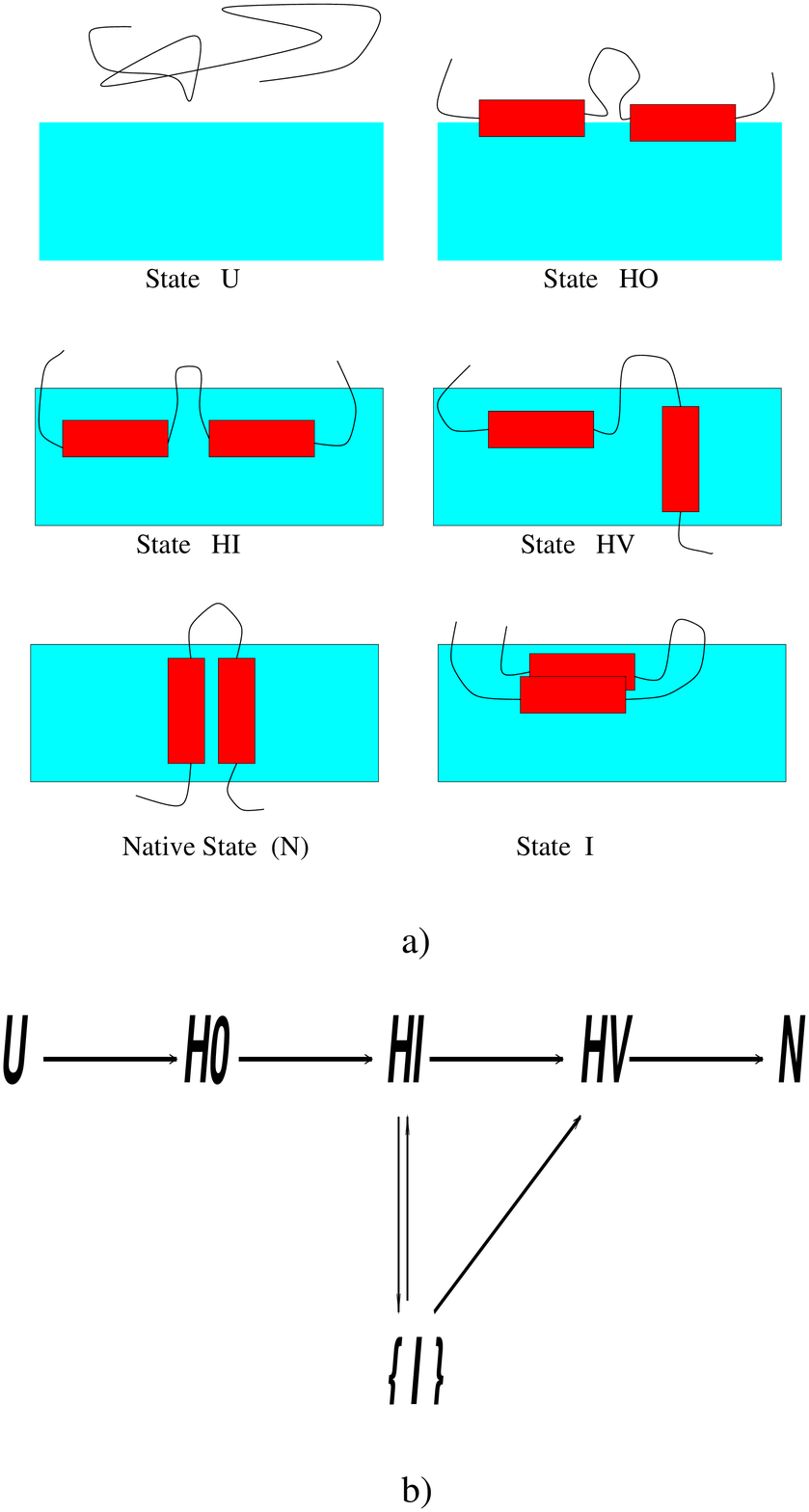}{0.6}
\caption{Seno}
\label{fig2}
\end{figure}

\newpage
\begin{figure}[tbp]
\inseps{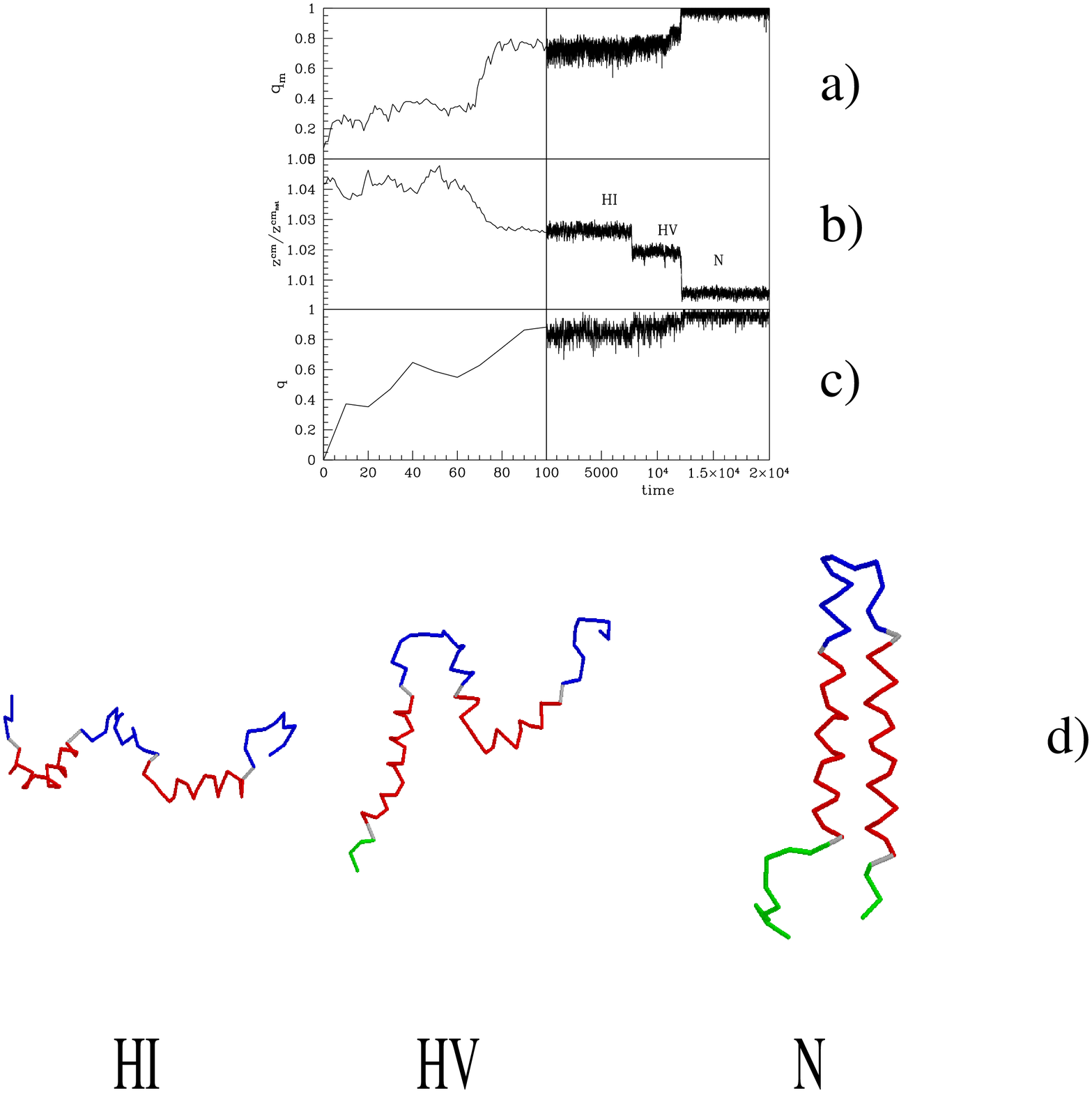}{0.7}
\caption{Seno}
\label{fig3}
\end{figure}

\newpage
\begin{figure}[tbp]
\inseps{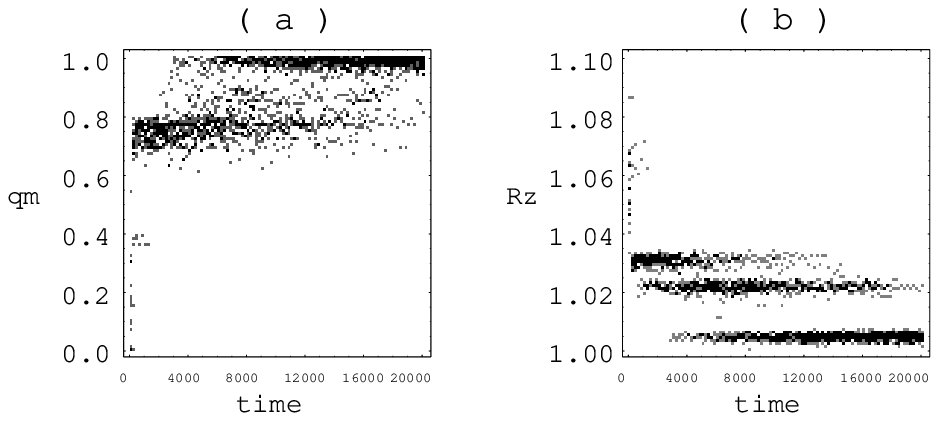}{1.5}
\caption{Seno}
\label{fig4}
\end{figure}

\newpage
\begin{figure}[tbp]
\inseps{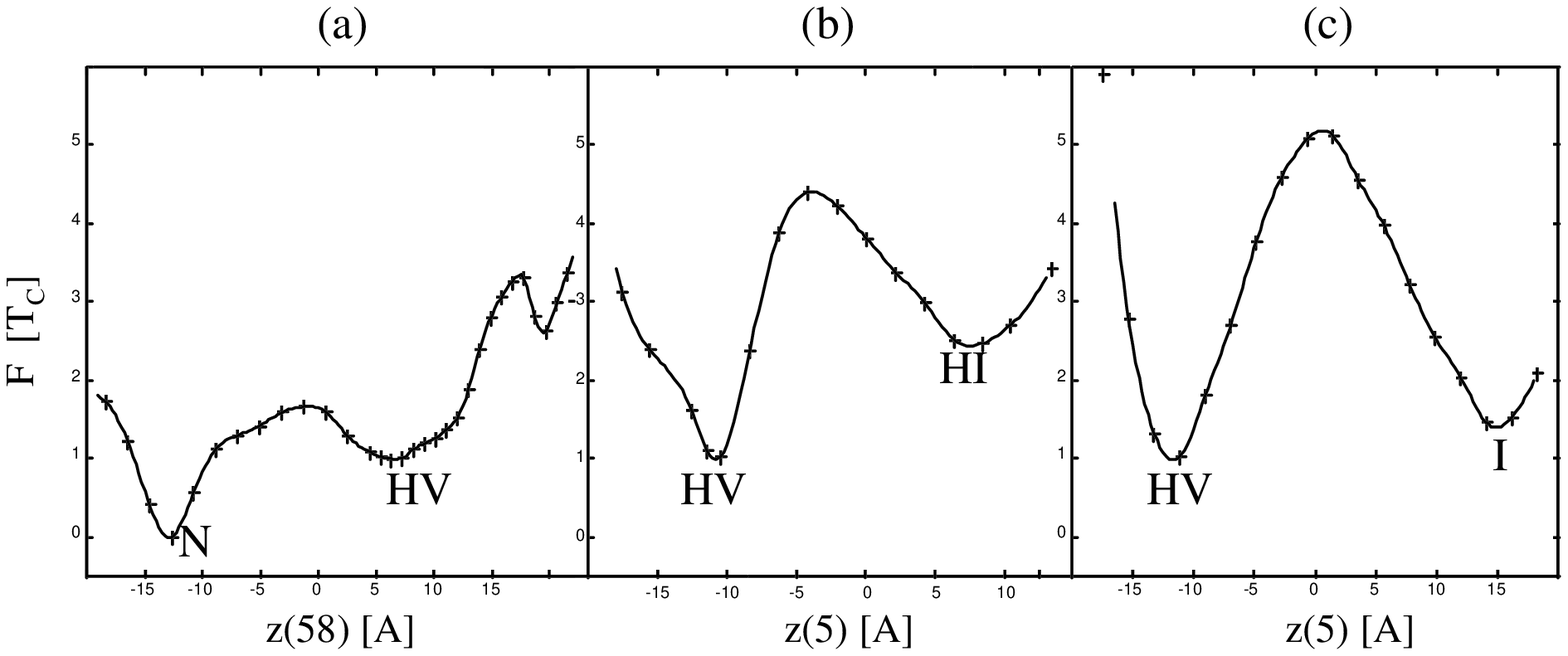}{0.7}
\caption{Seno}
\label{fig5}
\end{figure}


\begin{references}

\bibitem{editorial}  {\it Nature Structural Biology} {\bf 6}, 1-2 (1999).


\bibitem{Fersht3}  Fersht, A. R., W. H. Freeman, New York (1999).

\bibitem{Karplus}  Karplus, M. \& Sali, A. {\it Curr. Opin. Struct. Biol.} 
{\bf 5}, 58-73 (1995).


\bibitem{white1}  White, S. H. \& Wimley, W. C. {\it Ann. Rev. Biophys.
Biomol. Struct.} {\bf 28}, 319-365 (1999).



\bibitem{OM97}  Ostermeier, C. \& Michel, H. {\it Curr. Opin. Struct. Biol.} 
{\bf 7}, 697-701, (1997).

\bibitem{vH96}  von Heijne, G. {\it Prog. Biophys. Molec. Biol.} {\bf 66},
2, 113-139 (1996).

\bibitem{booth} Booth, P.J. {\it Folding \& Design}, R85-R92 (1997).

\bibitem{Biggin}  Biggin, P. C. \& Sansom, M. S. P. {\it Biophysical
Chemistry} {\bf 76}, 161-183 (1999).


\bibitem{goto}  Deber, C. M. \& Goto, N. K. {\it Nature Structural Biology} 
{\bf 3}, 815-818 (1996).


\bibitem{PE90}  Popot, J. L. \& Engelman, D. M. {\it Biochemistry} {\bf 29},
4031-4036 (1990).


\bibitem{Papu}  Pappu, R. V., Marshall, G. R. \& Ponder, J. W. {\it Nature
Structural Biology} {\bf 6}, 50-55 (1999).

\bibitem{MS1}  Milik, M. and Skolnick, J. {\em Proc. Natl. Acad. Sci. USA} 
{\bf 89}, 9391-9395, (1992)

\bibitem{MS2}  Milik, M. and J. Skolnick. {\em Proteins: Funct. Stru. Gen.} 
{\bf 15},10-25, (1993)


\bibitem{JW87}  Jacobs,R. E. \& ite, S. H., Biochemistry 26, 6127-6134 (1987)


\bibitem{Rose88}  Roseman, M. A., J. Mol. Biol. 200, 513-522 (1988).


\bibitem{preprint}  Wimley, S.C. \& White, S.H {\it Designing
Transmebrane $\alpha$-helices That Insert spontaneously} preprint
. University
of California, Irvine (2000)


\bibitem{ES81}  Engelman, D.M. \& Steitz, T.A. {\em Cell}, {\bf 23}, 411-422
(1981).

\bibitem{JW89}  Jacobs, R.E. \& White, S.H. {\em Biochemistry}, {\bf 28},
3421-3427 (1989).

\bibitem{JW86}  Jacobs, R.E. \& White, S.H. {\em Biochemistry}, {\bf 25},
2605-2612 (1986).

\bibitem{one}  Pinheiro, T. J. T. , El\"{o}ve, G.A., Watts, A. and Roder,
H.. {\em Biochemistry} {\bf 36},13122-13132 (1997)

\bibitem{two}  Rankin, S. E., Watts, A. and Pinheiro, T.J.T. {\em %
Biochemistry} {\bf 37}, 12588-12595 (1998)

\bibitem{three}  Bryson, E. A., Rankin, S.E., Carey,M. Watts, A. and
Pinheiro, T.J.T. {\em Biochemistry} {\bf 38} :9758-9767 (1999)




\bibitem{Fersht2}  Fersht, A. R. {\it Current Opinion in Struct. Biology} 
{\bf 5}, 79-84 (1995)

\bibitem{Fersht1}  Fersht, A. R. {\it Current Opinion in Struct. Biology} 
{\bf 7}, 3-9 (1997)

\bibitem{Micheletti2}  Micheletti, C. , Banavar, J. R., Maritan, A. \& Seno,
F. {\it Phys. Rev. Lett.} {\bf 82}, 3372-3375 (1999).

\bibitem{Eaton1}  Munoz, V. \& Eaton, W. A. {\it Proc. Natl. Acad. Sci.} 
{\bf 96}, 11311-11316 (1999).

\bibitem{Finkel}  Galzitskaya, O. V. \& Finkelstein, A. V. {\it Proc. Natl.
Acad. Sci.} {\bf 96}, 11299-11304 (1999).

\bibitem{Baker}  Alm, E. \& Baker, D. {\it Proc. Natl. Acad. Sci.} {\bf 96},
11305-11310 (1999).

\bibitem{7}  Chiti, F., Taddei, N., White, P. M., Bucciantini, M.,
Magherini, F., Stefani, M. and Dobson, C. M. {\it Nature Structural Biology} 
{\bf 6}, 1005, (1999).

\bibitem{8}  Martinez, J. C. and Serrano, L. {\it Nature Structural Biology} 
{\bf 6}, 1010, (1999).

\bibitem{9}  Riddle, D. S., Grantcharova, V. P., Santiago, J. V., Alm, E.,
Ruczinski, I. and Baker, D. {\it Nature Structural Biology} {\bf 6}, 1016,
(1999).

\bibitem{Woly}  Bryngelson, J. D. and Wolynes, P. G. {\it Proc. Nat. Acad.
Sci. USA} {\bf 84}, 7524, (1987).

\bibitem{KSE92}  Kahn, T. W., Sturtevant, J.M. and Engelman, D.M. {\em %
Biochemistry} {\bf 31},8829-8839 (1992)

\bibitem{KE92}  Kahn, T. W. \& Engelman, D.M.. {\em Biochemistry} {\bf 31}
,6144-6151 (1992)

\bibitem{Garel} Garel, T., Huse, D.A., Leibler, L. \& Orland, H. 
{\it Europhys. Lett.} {\bf 8}, 9-12  (1989)

\bibitem{Riva}
Maritan, A.,   Riva, M.P. \& Trovato, A.
{\it J. Phys. A: Math. Gen.} {\bf 32},  L275-L280 (1999)



\bibitem{Go}  Taketomi, H., Ueda, Y. \& Go, N. {\it Int. J. Pept. Protein
Res.} {\bf 7}, 445-459 (1975).

\bibitem{new1} Guo, Z. \& Thirumalai, D. {\it Journal of Molecular
Biology}
{\bf 263}, 323-343 (1996)


\bibitem{new2} Takada, S. , Luthey-Schulten, Z. \& Wolynes, P.G. {\it
Jour. of Chem. Phys.} {\bf 110}, 11616-11629 (1999)
 
\bibitem{Binder}  Geroff, I., Milchev, A., Binder, K. \& Paul, W. {\it J.
Chem. Phys.} {\bf 98}, 6256-6539 (1993).

\bibitem{TJOW}  M. C. Tesi, van Rensburg, E.J., Orlandini, E. \&
Whittington, S. G. {\it J. Stat. Phys.} {\bf 29}, 2451-2463 (1996).

\bibitem{FS89}  Ferrenberg, A. M. \& Swendsen, R. H. {\it Phys. Rev. Lett.} 
{\bf 63}, 1195-1198 (1989).
\end{references}
\end{document}